\documentclass[twoside]{article}
\usepackage{fleqn,espcrc2}

\usepackage{graphicx,psfrag,epsfig,epsf}
\usepackage[figuresright]{rotating}

\newcommand{\AmS}{{\protect\the\textfont2
  A\kern-.1667em\lower.5ex\hbox{M}\kern-.125emS}}

\title{
\begin{flushright}\normalsize
\vspace{-1cm}
{\tt LMU 11/04} \\ 
{\tt September 2004}
\end{flushright}
Non-leptonic $B$-decays, CP violation \& the UT\thanks{Invited talk at the 11th International Conference on {\it Quantum Chromodynamics}, Montpellier, 
France (5--10th July 2004)}}

\author{A. Salim Safir
\address{Ludwig-Maximilians-Universit\"at M\"unchen, Department f\"ur Physik, \\
Theresienstra\ss e 37, D-80333 Munich, Germany}}

\begin{document}

\date{{\tt LMU 26/03}}

\begin{abstract}
We study the implication of the time-dependent CP asymmetry in 
$B\to\pi^+\pi^-$ decays on the extraction of weak phases
taking into account the precise measurement of $\sin 2\beta$, obtained 
from the ``gold-plated''mode $B\to J/\psi K_S$.
Predictions and uncertainties for the hadronic parameters are investigated in QCD factorization. Furthermore, independent theoretical and experimental tests 
of the factorization framework are briefly discussed. Finally, a model-independent bound on the unitarity triangle from CP violation in $B\to\pi^+\pi^-$ and $B\to J/\psi K_S$ is derived.

\end{abstract}

\maketitle

\section{Introduction}
In the standard model (SM), the only source of CP violation 
is the Kobayashi-Maskawa phase \cite{CKM-KM}, localized in the Unitarity Triangle (UT) of the Cabibbo-Kobayashi-Maskawa (CKM) matrix \cite{CKM-KM,CKM-C}.
Thanks to the precise measurements at the current $B$-factories, CP violation  could be established in $B_d \to J/\psi K_S$~\cite{Aubert:2001nu,Abe:2001xe}, leading to a precise measurement of $\sin2\beta$, where the current world average yields~\cite{HFAG} $\sin 2\beta =0.739\pm 0.048$.
The extractions of the other two angles $\alpha$ and $\gamma$ are expected mainly through CP violation in the charmless $B$ decays, such as $B_d\to \pi\pi$ and similar modes~\cite{BABAR1}. The current $B$-factories measurements have been averaged to yield \cite{HFAG}:
\begin{eqnarray}\nonumber
S_{\pi\pi} =  -0.74\pm 0.16, \,\qquad C_{\pi\pi} =  -0.46\pm 0.13.
\end{eqnarray}
On the theoretical side, the analysis is challenging due to
the need to know the ratio of penguin-to-tree amplitude contributing 
to this process. In this talk, we present the result of \cite{safir,BS}, where a transparent 
method of exploring the UT through the CP violation in $B\to \pi^+\pi^-$, 
combined with the ``gold-plated'' mode $B_d \to J/\psi K_S$ has been proposed. A model independent lower bound on the CKM parameters as functions of $S_{\pi\pi}$ and 
$\sin 2\beta$ is derived. Our estimate of the hadronic parameters are carried out in QCD factorization (QCDF) and confronted to other approaches.

\section{Basic Formulas}
The time-dependent CP asymmetry in $B\to\pi^+\pi^-$ decays
is defined by
\begin{eqnarray}\label{acppipi}
A^{\pi\pi}_{CP}(t)
= - S_{\pi\pi}\, \sin(\Delta m_B t) + C_{\pi\pi}\, \cos(\Delta m_B t),
\end{eqnarray}
\begin{equation}\label{scxi}
{\rm where}\qquad
S_{\pi\pi}=\frac{2\, {\rm Im}\xi}{1+|\xi|^2},\,\,\,
C_{\pi\pi}=\frac{1-|\xi|^2}{1+|\xi|^2},\,\,\,
\end{equation}
with $\xi=e^{-2 i\beta}\,\frac{e^{-i\gamma}+P/T}{e^{+i\gamma}+P/T}$, and 
$\beta$ and $\gamma$ are CKM angles which are related to the Wolfenstein 
parameters $\bar\rho$ and $\bar\eta$ in the usual way \cite{Wolf}.

The penguin-to-tree ratio $P/T$ can be written as
$P/T=r e^{i\phi}/\sqrt{\bar\rho^2+\bar\eta^2}\equiv r e^{i\phi}/R_b$.
The real parameters $r$ and $\phi$ defined in this way are
pure strong interaction quantities without further dependence
on CKM variables.

For any given values of $r$ and $\phi$ a measurement of $S_{\pi\pi}$ and
$C_{\pi\pi}$ defines a curve in the ($\bar\rho$, $\bar\eta$)-plane,   
expressed respectively through 
\begin{eqnarray}\label{srhoeta}
S_{\pi\pi}=\frac{2\bar\eta [R_b^2-r^2-\bar\rho(1-r^2)+
       (R_b^2-1)r \cos\phi]}{((1-\bar\rho)^2+\bar\eta^2)
         (R_b^2+r^2 +2 r\bar\rho \cos\phi)}
\end{eqnarray}
\begin{eqnarray}
\label{crhoeta}
{\rm and }\qquad
C_{\pi\pi}=\frac{2 r\bar\eta\, \sin\phi}{
   R_b^2+r^2 +2 r\bar\rho \cos\phi}.
\end{eqnarray}

The penguin parameter $r\, e^{i\phi}$ has been computed
in \cite{BBNS1} in the framework of QCDF.
The result can be expressed in the form
\begin{equation}\label{rqcd}
r\, e^{i\phi}= -
\frac{a^c_4 + r^\pi_\chi a^c_6 + r_A[b_3+2 b_4]}{
 a_1+a^u_4 + r^\pi_\chi a^u_6 + r_A[b_1+b_3+2 b_4]},
\end{equation}
where we neglected the very small effects from electroweak
penguin operators. A recent analysis gives \cite{safir,BS}
\begin{equation}\label{rphi}
r=0.107\pm 0.031, \qquad \phi=0.15\pm 0.25,
\end{equation}
where the error includes an estimate of potentially important
power corrections.
In order to obtain additional insight into the structure of
hadronic $B$-decay amplitudes, it will be also interesting to extract these 
quantities from other $B$-channels, or using other methods.
In this perspective, we have considered them in a simultaneous expansion 
in $1/m_b$ and $1/N_C$ ($N_C$ is the number of colours) in (\ref{rqcd}). 
Expanding these coefficients to first order in $1/m_b$ and
$1/N_C$ we find that
the uncalculable power corrections $b_i$ and $H_{\pi\pi,3}$ do
 not appear in (\ref{rqcd}), to
which they only contribute at order $1/m_b N_C$. Using our default input parameters, one obtains the central value \cite{safir}: 
$(r_{N_C},\phi_{N_C})=(0.084,0.065)$,
which seems to be in a good agreement with the standard QCDF framework at the next-to-leading order.

As a second cross-check, one can extract
$r$ and $\phi$ from $B^+\to\pi^+\pi^0$ and  $B^+\to\pi^+ K^0$, leading to the central value \cite{safir} $(r_{SU3},\phi_{SU3})=(0.081, 0.17)$, in agreement with the above results\footnote{one can compare also $r_{SU3}$ to its experimental value $r_{SU3}^{exp}=0.099\pm0.014$.}, although their definitions differ slightly from $(r,\phi)$ (see \cite{safir} for further discussions). 
\section{UT through CP violation observables}
It is possible to fix the UT by combining the information from $S_{\pi\pi}$
with the value of $\sin 2\beta$, well known from the 
``gold-plated'' mode $B\to J/\Psi K_S$.
\begin{figure}[t!]
\psfrag{S}{$S_{\pi\pi}$}
\psfrag{etabar}{\hspace*{0.5cm}$\bar\eta$}
\begin{center}
\epsfig{file=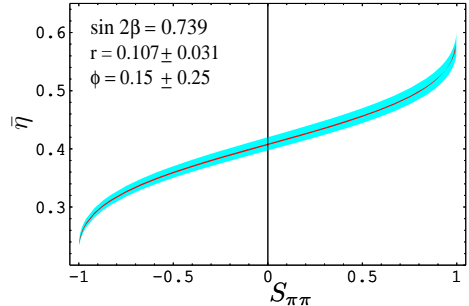,width=6cm,height=4cm}
\end{center}
\vspace{-1.cm}
{\caption{\it CKM phase $\bar\eta$ as a function of 
$S_{\pi\pi}$.
The dark (light) band reflects the theoretical uncertainty
in the parameter  $\phi$ ($r$).      
\label{fig:etabspp}}}
\end{figure}
The angle $\beta$ of the UT is given by
\begin{equation}\label{taus2b}
\tau\equiv\cot\beta=\sin 2 \beta\Bigg(1-\sqrt{1-\sin^2 2\beta}\Bigg)^{-1}.
\end{equation}
The current world average \cite{HFAG} 
$\sin 2\beta =0.739\pm 0.048$, implies
$\tau=2.26\pm 0.22.$
Given a value of $\tau$, $\bar\rho$ is related to $\bar\eta$
by
$\bar\rho = 1-\tau\, \bar\eta$.
The parameter $\bar\rho$ may thus be eliminated from $S_{\pi\pi}$
in (\ref{srhoeta}), which can be solved for $\bar\eta$ to yield
\begin{eqnarray}\label{etataus}
\bar\eta =\frac{1}
{(1+\tau^2)S_{\pi\pi}} 
\Bigg[{\tilde S}(1+r \cos\phi) \\
\hspace*{0cm}-\sqrt{(1-S_{\pi\pi}^2)(1+r^2+2 r\cos\phi)-
  {\tilde S}^2 r^2 \sin^2\phi}\Bigg],\nonumber 
\end{eqnarray}
with ${\tilde S}=(1+\tau S_{\pi\pi})$
The two observables $\tau$ (or $\sin 2\beta$) and $S_{\pi\pi}$ determine
$\bar\eta$ and $\bar\rho$ once the theoretical penguin parameters
$r$ and $\phi$ are provided.

The determination of $\bar\eta$ as a function of $S_{\pi\pi}$ is shown
in Fig. \ref{fig:etabspp}, which displays the theoretical uncertainty 
from the penguin parameters $r$ and $\phi$ in QCDF. Since 
the dependence on $\phi$ enters in (\ref{etataus}) only at second order, 
it turns out that its sensitivity is rather mild in contrast to $r$. 
In the determination of $\bar\eta$ and $\bar\rho$ described
here discrete ambiguities do in principle arise, however they are ruled out using the standard fit of the UT (see \cite{safir} for further discussions).

After considering the implications of $S_{\pi\pi}$ on the UT, let's explore now $C_{\pi\pi}$.
Since $C_{\pi\pi}$ is an odd function of $\phi$,
it is therefore sufficient to restrict the discussion to positive values
of $\phi$. A positive phase $\phi$ is obtained by the perturbative
estimate in QCDF, neglecting soft phases with power
suppression. For positive $\phi$ also $C_{\pi\pi}$ will be positive,
assuming $\bar\eta > 0$, and a sign change in $\phi$ will simply
flip the sign of $C_{\pi\pi}$.

In contrast to the case of $S_{\pi\pi}$, the hadronic quantities $r$ and $\phi$
play a prominent role for $C_{\pi\pi}$, as can be seen in (\ref{crhoeta}).
This will in general complicate the interpretation of an experimental
result for $C_{\pi\pi}$.

\begin{figure}[t]
\psfrag{phi}{\hspace*{-0.2cm} $\phi$}
\psfrag{r}{$r$}
\begin{center}
\epsfig{file=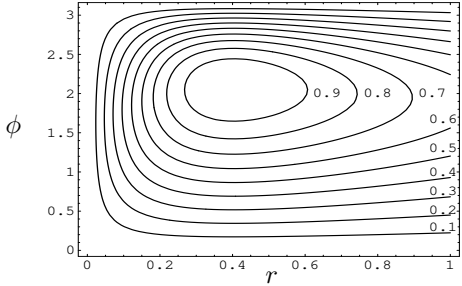,width=6cm,height=4cm}
\end{center}
\vspace{-1.cm}
{\caption{\it Contours of constant $C_{\pi\pi}$ in the ($r$, $\phi$)-plane
for the SM best-fit result $(\bar\rho,\bar\eta)=(0.20,0.35)$ 
\cite{CKMF}. \label{fig:cpipi}}}
\end{figure}

The analysis of $C_{\pi\pi}$ becomes more transparent if we fix the
weak parameters and study the impact of $r$ and $\phi$.
An important application is a test of the SM,
obtained by taking $\bar\rho$ and $\bar\eta$ from a
SM fit and comparing the experimental result for $C_{\pi\pi}$
with the theoretical expression as a function of $r$ and $\phi$.
In Fig. \ref{fig:cpipi}, a useful representation is obtained by 
plotting contours of constant
$C_{\pi\pi}$ in the ($r$, $\phi$)-plane, for given values of $\bar\rho$ and $\bar\eta$.
Within the SM this illustrates the correlations between 
the parameters $(r,\phi)$ and observable $C_{\pi\pi}$. 

As it has been shown in \cite{safir}, a bound on the parameter $C_{\pi\pi}$ exists, given by
\begin{equation}\label{barc}
C_{max}=\frac{2 \kappa\, \sin\phi}{
  \sqrt{(1+\kappa^2)^2 -4 \kappa^2 \, \cos^2\phi}},
\end{equation}
with $\kappa\equiv r/R_b$ and where the maximum occurs at $\cos\gamma=-2\kappa\cos\phi/(1+\kappa^2)$.
If $\kappa=1$, 
no useful upper bound is obtained.
However, if $\kappa < 1$, then $C_{max}$ is maximized for
$\phi=\pi/2$, yielding the general bound
$C < \frac{2\kappa}{1+\kappa^2}$.
For the conservative bound $r < 0.15$, $\kappa < 0.38$ this implies
$C_{\pi\pi} < 0.66$. The bound on $C_{\pi\pi}$ can be strengthened by using information
on $\phi$, as well as on $\kappa$, and employing (\ref{barc}).
Then $\kappa < 0.38$ and $\phi < 0.5$ gives $C_{\pi\pi} < 0.39$.
\section{Model Independent bound on the UT}
As has been shown in \cite{BS}, the following inequality
can be derived from (\ref{etataus}) for $-\sin 2\beta\leq S_{\pi\pi}\leq 1$
\begin{equation}\label{etb3}
\bar\eta\geq\frac{1+\tau S_{\pi\pi}-\sqrt{1-S_{\pi\pi}^2}}{(1+\tau^2)S_{\pi\pi}}(1+r\cos\phi).
\end{equation}
This bound is still {\it exact} and requires no information
on the phase $\phi$. 

Assuming now $-90^\circ\leq\phi\leq 90^\circ$, we have $1+r\cos\phi \geq 1$ and
\begin{equation}\label{etabound}
\bar\eta\geq\frac{1+\tau S_{\pi\pi}-\sqrt{1-S_{\pi\pi}^2}}{(1+\tau^2)S_{\pi\pi}}.
\end{equation}
We emphasize that this lower bound on $\bar\eta$ depends only on the
observables $\tau$ and $S_{\pi\pi}$ and is essentially free of hadronic
uncertainties. 
Since both $r$ and $\phi$ are expected to be
quite small, we anticipate that the lower limit (\ref{etabound})
is a fairly strong bound, close to the actual value of $\bar\eta$ itself
(see \cite{safir} for further details).
We also note that the lower bound (\ref{etabound}) represents the
solution for the unitarity triangle in the limit of vanishing
penguin amplitude, $r=0$. In other words, the model-independent
bounds for $\bar\eta$ and $\bar\rho$ are simply obtained
by ignoring penguins and taking $S_{\pi\pi}\equiv\sin 2\alpha$ when fixing
the unitarity triangle from $S_{\pi\pi}$ and $\sin 2\beta$.
\begin{figure}[t]
\psfrag{S}{$S_{\pi\pi}$}
\psfrag{etab}{$\bar\eta$}
\begin{center}
\epsfig{file=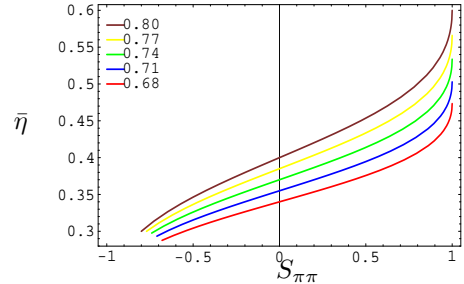,width=6cm,height=4cm}
\end{center}
\vspace{-1.cm}
{\caption{\it Lower bound on $\bar\eta$ as a function
of $S_{\pi\pi}$ for various values of $\sin 2\beta$
.  \label{fig:etabound}}}
\end{figure}
Let us briefly comment on the second solution for $\bar\eta$, which has 
the minus sign in front of the square root in (\ref{etataus}) replaced by a
plus sign. For positive $S_{\pi\pi}$ this solution is always larger than
(\ref{etataus}) and the bound (\ref{etabound}) is unaffected.
For $-\sin 2\beta\leq S_{\pi\pi}\leq 0$ the second solution gives a negative
$\bar\eta$, which is excluded by independent information on the
UT (for instance from $\varepsilon_K$). 

Because we have fixed the angle $\beta$, or $\tau$, the lower bound
on $\bar\eta$ is equivalent to an upper bound on $\bar\rho=1-\tau\bar\eta$. 
The constraint (\ref{etabound}) may also be expressed as a lower bound
on the angle $\gamma$
or a lower bound on $R_t$ (see \cite{safir} for further details).
In Figs. \ref{fig:etabound},
we represent the lower bound on
$\bar\eta$  as a function of $S_{\pi\pi}$ for various values of
$\sin 2\beta$. From Fig. \ref{fig:etabound} we observe that 
the lower bound on $\bar\eta$ becomes stronger as either
$S_{\pi\pi}$ or  $\sin 2\beta$ increase. 

\begin{figure}[t]
\psfrag{etab}{$\bar\eta$}
\psfrag{rob}{$\bar\rho$}
\psfrag{sin}{\small $\sin 2\beta=0.739\pm 0.048$}
\begin{center}
\epsfig{file=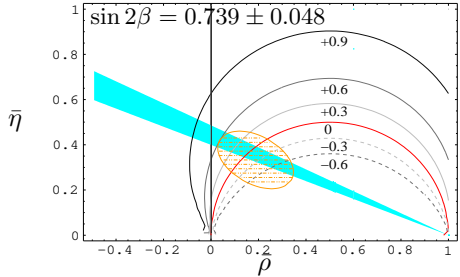,width=6cm,height=5cm}
\vspace{-2.cm}
{\caption{\it 
 Model-independent bound on the $(\bar\rho,\bar\eta)$ plane from 
$\sin 2\beta$ (shaded area) and $S_{\pi\pi}$. 
The result of a standard UT fit (dotted ellipse,
from \cite{CKMF}) is overlaid for comparison.   
\label{fig:utbound}}}
\end{center}
\end{figure}

In Fig. \ref{fig:utbound} we illustrate the region in the
$(\bar\rho,\bar\eta)$ plane that can be constrained by the
measurement of $\sin 2\beta$ and $S_{\pi\pi}$ using the bound
in (\ref{etabound}).
We finally note that the condition $r\cos\phi > 0$,
which is crucial for the bound, could be independently checked
\cite{SKK} by measuring the mixing-induced CP-asym\-me\-try in
$B_s\to K^+ K^-$, the U-spin counterpart of the $B_d\to\pi^+\pi^-$ mode \cite{DF}.
%
\section{Summary}
In this talk, we have proposed strategies to extract
information on weak phases from CP violation
observables in $B\to\pi^+\pi^-$ decays even in the presence
of hadronic contributions related to penguin amplitudes.
Assuming knowledge of the penguin pollution, an efficient use of mixing-induced CP violation in $B\to\pi^+\pi^-$
decays, measured by $S_{\pi\pi}$, can be made by combining it with the corresponding
observable from $B\to J/\psi K_S$, $\sin 2\beta$, to obtain the unitarity triangle parameters $\bar\rho$ and $\bar\eta$.
The sensitivity on the hadronic quantities, which have typical
values $r\approx 0.1$, $\phi\approx 0.2$, is very weak.
In particular, there are no first-order corrections in $\phi$.
For moderate values of $\phi$ its effect is negligible.

Concerning our penguin parameters, 
namely $r$ and $\phi$, they were 
investigated systematically within the QCDF framework.
To validate our theoretical predictions, we have calculate these parameters in the $1/m_b$ and $1/N_C$ expansion, which exhibits a good framework to control the uncalculable power corrections, in the factorization formalism. As an alternative proposition,  we have also considered to extract $r$ and $\phi$ from other $B$ decay channels, such as $B^+\to\pi^+\pi^0$ and $B^+\to\pi^+ K^0$, relying on the SU(3) argument. Using these three different approaches, we found a compatible picture in estimating these hadronic parameters.
%
\section*{Acknowledgements}
I thank the organizers for their invitation and I am very grateful to Gerhard Buchalla for the extremely pleasant collaboration. This work is supported by the 
DFG under contract BU 1391/1-2.

\end{document}